\documentclass[12pt]{iopart}

\usepackage{graphicx}

\begin{document}

\title[Middelmann et al.: Long range transport in a far-detuned lattice]{Long range transport of ultra cold atoms in a far-detuned 1D optical lattice}

\author{Thomas Middelmann, Stephan Falke, Christian Lisdat, and Uwe Sterr}

\address{Physikalisch-Technische Bundesanstalt, Bundesallee 100,
38116 Braunschweig, Germany}
\ead{thomas.middelmann@ptb.de}
\begin{abstract}
We present a novel method to transport ultra cold atoms in a focused optical lattice over macroscopic distances of many Rayleigh ranges. With this method ultra cold atoms were transported over 5~cm in 250~ms without significant atom loss or heating. By translating the interference pattern together with the beam geometry the trap parameters are maintained over the full transport range. Thus, the presented method is well suited for tightly focused optical lattices that have sufficient trap depth only close to the focus. Tight focusing is usually required for far-detuned optical traps or traps that require high laser intensity for other reasons. The transport time is short and thus compatible with the operation of an optical lattice clock in which atoms are probed in a well designed environment spatially separated from the preparation and detection region.
\end{abstract}

\maketitle

\section{Introduction}
The transport of trapped ions \cite{kie02, stu11}, ultra cold atoms \cite{gre01a} or even quantum degenerate gases \cite{gus02, lea02, kle08} is a well developed tool in atomic physics and quantum optics, which is applied in macroscopic traps or even in chip traps \cite{hae01a, kru03a, hom05}. The typical aim is to bring an atomic sample into a specific environment that is spatially separated from the preparation region of the sample and provides dedicated conditions required for further experiments. 
In many cases in this environment vacuum conditions, magnetic or electric field conditions are different, or optical access is more favourable \cite{nak05, lew03, sch07g}; bringing atoms into controlled interaction with surfaces \cite{obr07a, mar11a} or other particles is also of interest. For neutral atoms, long range transport is very often realized by moving magnetic traps or hybrid optical-magnetic traps \cite{gri09}. However, magnetic trapping is not applicable for non-magnetic atoms as e.g. alkaline earth like atoms in their ground states, or mixtures of atoms in different magnetic sublevels \cite{ste98}.

If only optical traps are applied, often a compromise between trap parameters, transport speed, and transport distance has to be found. 
With focused dipole traps (optical tweezers) long range transport has been realized by moving the optics \cite{gus02}, but the possible transport speed is limited due to the weak confinement.  In optical lattices the confinement in beam direction is much stronger, allowing much higher accelerations. In near-resonant lattices, where no focusing is needed to obtain sufficiently deep traps with the available laser power, long range transport has been realized utilizing frequency differences between the counter propagating beams \cite{pei97}. The restriction to near resonant traps is similar in lattices formed from Bessel beams \cite{sch06a} because only a part of the laser power per beam is contributing to the local trap depth.

In this paper we describe the transport of atoms in a novel moving lattice setup that combines the advantages of moving optics and transport in an optical lattice to allow for transport of ultra cold atoms over long distances in short time. We move all optics of the optical lattice setup simultaneously with two air bearing translation stages that allow for fast transport and high acceleration while providing extremely smooth motion (Fig.~\ref{fig:setup}). 
With this method the transport can be integrated into the cycle of an optical lattice clock (in our case with strontium \cite{fal11}) that requires a tightly focused, far-detuned optical lattice operated at the magic wavelength \cite{kat03, wes11}. Since the transport time is short the method is also compatible with high stability clock operation.
The moving lattice setup (Fig.~\ref{fig:setup}) enables investigation of the atoms in a region with little optical access, as e.g. for a dc Stark shift measurement in a precision capacitor or a suppression of room temperature blackbody radiation in a cryogenic environment. As the influence of room temperature blackbody radiation on the clock frequency is currently limiting the fractional uncertainty of strontium optical lattice clocks to about $1\times 10^{-16}$ \cite{cam08b, fal11}, an interrogation of the atoms in a liquid nitrogen cooled environment would reduce the blackbody shift and its uncertainty to a few times $10^{-18}$ \cite{mid11},which is a strong motivation to follow this path of investigation.
The moving lattice setup and the experimental apparatus is described in the following section. In Sec.~\ref{sec:trans} we describe systematic investigations of the transport, followed by a discussion of the implementation of the moving lattice into the clock cycle in Sec.~\ref{sec:clock}.

%
%%%%%%%%%%%%%%%%%%%%%%%%%%%%%%%%%%%%%%%%%%%%%%%%%%%%%%%%%%%%%%%%%%%%%%%
%
%		Experimental Setup
%
%%%%%%%%%%%%%%%%%%%%%%%%%%%%%%%%%%%%%%%%%%%%%%%%%%%%%%%%%%%%%%%%%%%%%%%
%
\section{Experimental setup}\label{sec:setup}

\begin{figure}[!t]
\centering
\includegraphics[width=\columnwidth]{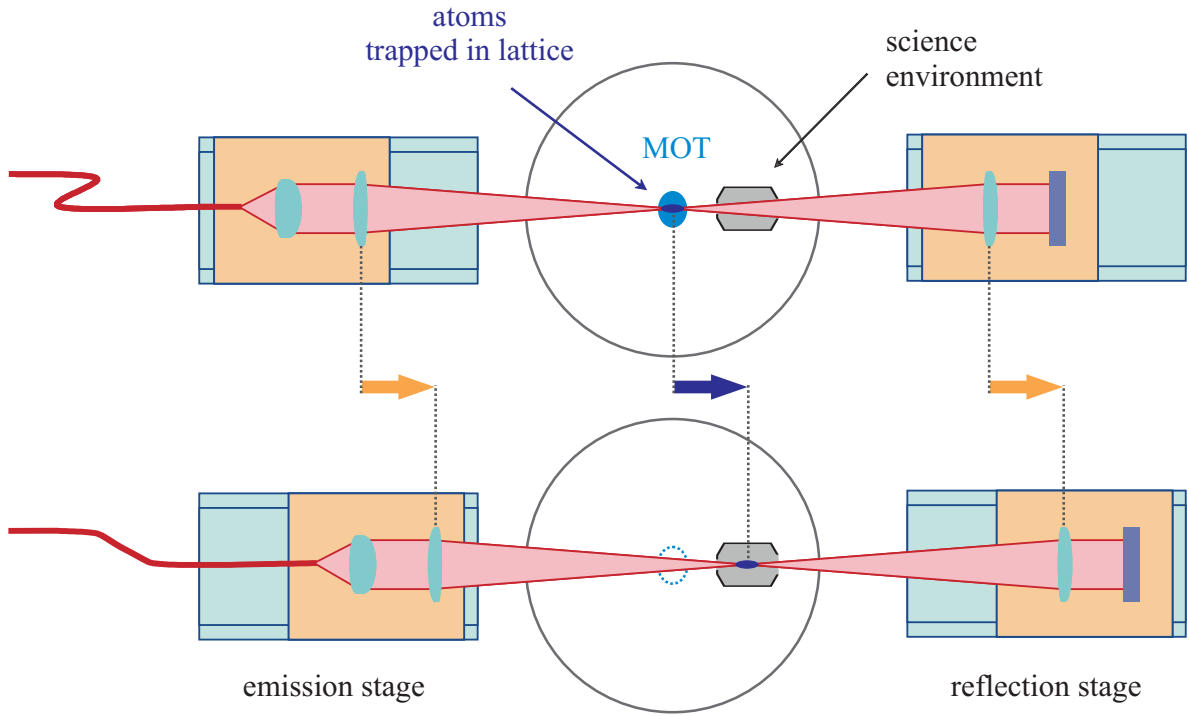}
\caption{Sketch of the moving lattice setup. Atoms are loaded from the magneto-optical trap (MOT) into the lattice (upper part) and transported to the 
``science environment'' by moving the lattice optics on both sides of the vacuum chamber with air bearing translation stages (lower part).}
\label{fig:setup}
\end{figure}

In the experiments presented here we work with ultra cold $^{88}$Sr atoms. 
A thermal beam of atoms is produced from an oven at a temperature $T\approx$~500~$^\circ$C. A collimated beam is defined by a pinhole downstream that also acts as differential pumping stage between oven and main chamber. 
The atoms are decelerated in a Zeeman slower. Leaving the slower, the atoms are collimated and deflected by an optical molasses towards the magneto optical trap (MOT). For deceleration, deflection, and trapping, light red-detuned from the 461~nm transition $^1$S$_0$ -- $^1$P$_1$ is used. To decrease atom losses during the 461~nm MOT phase we apply repump light on the transitions $^3$P$_0$ -- $^3$S$_1$ (679~nm) and $^3$P$_2$ -- $^3$S$_1$ (707~nm). Within a loading time of 20~ms we can trap $2\times 10^{6}$ atoms at 5~mK; at maximum $6\times 10^{7}$ atoms can be loaded into the MOT within 800~ms. 

To further cool the atoms a second MOT phase is applied, operating on the narrower 689~nm intercombination transition $^1$S$_0$ -- $^3$P$_1$. Since the natural linewidth is small ($\approx$~7~kHz) the light is frequency modulated at 30~kHz with a peak-to-peak excursion of 3~MHz to increase the velocity capture range. After 90~ms the frequency modulation is turned off and single frequency light with decreased intensity is applied for 50~ms. This results in a temperature of the strontium cloud of a few~$\mu$K.

During this last cooling phase more than 25~\% of the atoms are loaded into the horizontally oriented 1D optical lattice that crosses the MOT. To create the optical lattice a collimated laser beam is focused on the MOT  with an achromatic lens ($f=400$~mm) to a waist radius $w_0 = 65$~$\mu$m, collimated with a second lens of the same focal length and retro-reflected to obtain the desired interference pattern in which the atoms are trapped. 
The lattice light wavelength of 813~nm is magic for the strontium clock transition \cite{kat03,wes11}, i.e., both clock states are shifted by same amount and the trap parameters are the same.
This light is provided by a Ti:Sapphire laser coupled to a polarization-maintaining large mode area fibre. A power of up to 600~mW is available at the fibre output. Residual reflections on the windows of the vacuum chamber and lenses lead to a reduction of trap and modulation depth (by imbalanced intensity in both beams) to $\approx$22~$\mu$K. Since the lattice is oriented horizontally, the trapping potential in downward direction is reduced by $\approx$9~$\mu$K due to gravity. In the potential minima the trap frequencies are about 79~kHz and 220~Hz in longitudinal and radial direction respectively.

The optics of the lattice setup on both sides of the vacuum chamber are mounted on two air bearing translation stages (Fig. \ref{fig:setup}). A first test of the optical setup alone was presented in \cite{mid11}, for completeness we recall the important parameters here. The commercial stages allow for 100~mm travel, a top speed of 300~mm/s, and a maximum acceleration of 10~m/s$^2$. The fibre tip, collimation lens, polarization optics and the lattice lens are mounted on the ``emission stage'' on one side of the vacuum system. On the opposing side the second lattice lens and the retro-reflecting mirror are mounted on the ``reflection stage''. To maintain the interference pattern during transport, the two waists need to stay overlapped. Thus the mechanical quality of the motion is crucial because already a radial displacement of half a beam waist (here 33~$\mu$m) reduces the lattice depth to about 77~\%. 
 
To ensure a good overlap of the two waists a straight motion of the stages with low dynamic and static tilting and a good mutual alignment of the two translation stages is crucial. 
Hence we are using synchronized air bearing translation stages mounted on adjustment frames each allowing for adjustment of four degrees of freedom: two transverse translations and two angular alignments to adjust the pointing of the translation axis. After adjustment, the frames can be rigidly locked. 
The mutual alignment of motional directions is better than 0.1~mrad, the mutual static tilting is less than $\mathrm{\pm15\: \mu rad}$ over the full travel path of 100~mm and the dynamic tilting of each stage at maximum acceleration is less than $\pm16$~$\mu$rad (pitch) and $\pm5$~$\mu$rad (yaw) as determined in the test setup \cite{mid11}. The dynamic pitch is roughly proportional to the acceleration. As an indicator of the overlap of the two lattice waists, we observe the retro-reflected lattice light, which is coupled back into the fibre. On the other side of the fibre it is extracted by a Faraday isolator and its power is recorded with a photo diode. If the lattice is well aligned the retro-reflected signal is maximized at the central position and drops at the end positions by 2~\%, which corresponds to a displacement of the two lattice waists by 10~$\mu$m. Additionally, during acceleration and deceleration, we observe a drop due to the dynamic pitch (Fig.~\ref{fig:traj} and Sec.~\ref{sec:trans}). 

The clock laser light is delivered via a polarization-maintaining optical fibre from an extended cavity diode laser in Littman configuration. Its frequency is pre-stabilised to an ultra-stable resonator made of ultra-low expansion glass (ULE) \cite{leg09, vog11}. A linewidth of below 1~Hz and a short term stability of $2\times10^{-15}$ is achieved. Tuneability and the possibility to pulse the light onto the atoms is provided by offset and switching acousto-optical modulators (AOMs). To ensure a constant phase of the clock laser field at the position of the atoms in the case of perturbations on the fibre or residual lattice movements, we stabilise the path length between the clock laser setup and the lattice mirror \cite{fal11a}, 
which maintains a constant distance to the interference pattern and thus to the position of the atoms.

To induce a nonzero dipole matrix element on the $^1$S$_0$ -- $^3$P$_0$ clock transition at 698~nm of $^{88}$Sr \cite{tai06}, we apply a homogeneous magnetic field of up to 3~mT. For spectroscopy we overlap the clock laser beam with the lattice through the lattice mirror (which is highly transparent for the radiation of the clock laser) and align it parallel to the lattice beams, ensuring interrogation of the atoms along the lattice axis with strong confinement and thus a small Lamb-Dicke-parameter. To probe the clock transition, Rabi spectroscopy of the atoms is performed with pulses between 1~ms and 90~ms duration. We detect the number of atoms remaining in the $^1$S$_0$ ground state after the excitation on the clock transition by monitoring fluorescence induced by light resonant with the  $^1$S$_0$ -- $^1$P$_1$ transition. The atoms are removed by the radiation pressure; then the atoms in the $^3$P$_0$ state are repumped to the ground state and are detected by a second fluorescence detection phase.

If spectroscopy is performed with transport into the science environment the transportation phases are inserted directly before and after the clock laser pulse. 
%
%%%%%%%%%%%%%%%%%%%%%%%%%%%%%%%%%%%%%%%%%%%%%%%%%%%%%%%%%%%%%%%%%%%%%%%
%
%		TRANSPORT
%
%%%%%%%%%%%%%%%%%%%%%%%%%%%%%%%%%%%%%%%%%%%%%%%%%%%%%%%%%%%%%%%%%%%%%%%
%
\section{Transport} \label{sec:trans}
\begin{figure}[!t]
\centering
\includegraphics[width=7cm]{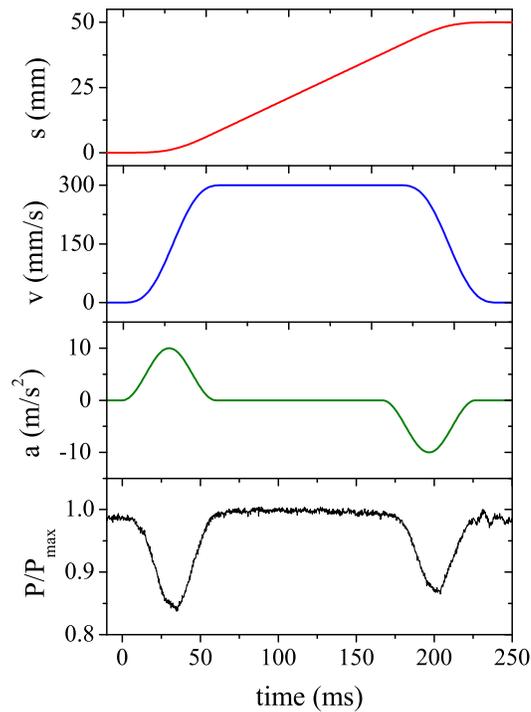}
\caption{Moving lattice trajectory to exhaust the capability of the air bearing translation stages and minimize jerks by a sinusoidal acceleration profile. A 50-mm-transport is completed in 230~ms. From top to bottom position, velocity, acceleration, and the normalized signal of the retro-reflected lattice light are shown.}
\label{fig:traj}
\end{figure}

We optimized the motion of the translation stages for quickness of the transport while minimizing the jerk to avoid temporary misalignment by inertial forces. In contrast to Ref.~\cite{mid11} we chose a sinusoidal acceleration profile
\begin{equation}
a(t) =  \pm a_{\rm max} \sin[\pi (t - t_0)/t_{\rm ramp} ]^2  \label{eq:accel}
\end{equation}
with maximum acceleration $a_{\rm max}=10$~m/s$^2$ and duration of the acceleration phase $t_{\rm ramp}=60$~ms to reach the maximum velocity of the stages of $v_{\rm max}=300$~mm/s. This velocity is maintained as long as necessary for the desired transport distance. The stages are stopped by an equivalent deceleration profile. A typical trajectory is presented in Fig.~\ref{fig:traj}, where a transport of 50~mm is completed in 230~ms. 

During motion the power of the retro-reflected light stays above 84~\% of the power at rest (bottom plot in Fig.~\ref{fig:traj}). This corresponds to a radial displacement of the lattice waists of about $0.4~w_{\rm 0}$. For displacements smaller than the beam waist $w_0$ the trap depth reduction is proportional to the drop of the signal. In addition to this trap depth reduction related to accelerations of the air bearing stages, one may loose trap depth because the overlap of the interfering laser beams is reduced once the stages are at a different position. Using again the retro-reflected signal, we determine a position dependent change in trap depth smaller than 2~\%.

\begin{figure}[!t]
\centering
\includegraphics[width=\columnwidth]{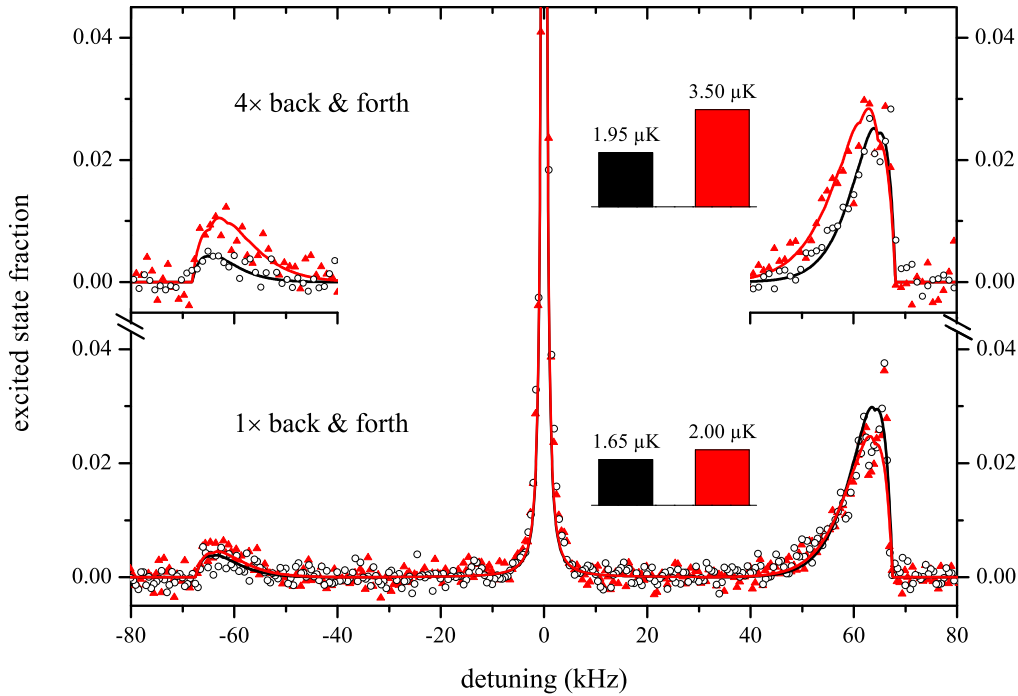}
\caption{Sideband spectra after transporting the atoms once (main graph) and four times (insets) back and forth for 50~mm (solid red triangles) and after holding the atoms in the static lattice for the same time (open black circles). The solid curves are combinations of fits to the carrier and the sidebands, the latter according to \cite{bla09a}.}
\label{fig:sidebands}
\end{figure}

From these data an efficient transport of atoms is expected. 
We evaluate the performance of the transport by loading atoms into the lattice and moving them back and forth over 5~cm. Their number is compared to the atom number in a cycle without transport using a hold time equal to the transport time. We achieve an efficiency of 96~\% for a complete roundtrip. 
The quality of the beam overlap is confirmed by measuring the frequencies of the longitudinal sidebands of the clock transition (Fig.~\ref{fig:sidebands}) at different translation stage positions. No variation larger than 1.5~\% is found, in agreement with the optical investigation.

To investigate heating by the transport we use the longitudinal sideband spectra to extract the atomic temperature \cite{bla09a}. Spectra are recorded after transporting the atoms for one or for four round trips and after holding them for the same time in the static lattice (Fig.~\ref{fig:sidebands}). Each round trip corresponds to a transport of 50~mm back and forth, using the trajectory presented in Fig.~\ref{fig:traj}. We add a delay of 0.6~s at each turning point of the trajectories to avoid resonant vibrations of the frames and overload of the current controller of the stages.

We observe the weak longitudinal sideband spectra with a clock pulse of 1.5~s duration at a peak intensity of the clock laser of 2~mW/cm$^2$ and a magnetic field of 3~mT. The four-round-trip spectra in Fig.~\ref{fig:sidebands} are each the average of four measurements, as the signal to noise ratio is decreased due to losses during the long trapping time (the lattice lifetime is $\approx$4.4~s). We determine the longitudinal temperature from the ratio of red and blue sideband area \cite{bla09a}, and obtain in the one-round-trip case 1.65~$\mu$K and  2~$\mu$K without and with transport respectively and 1.95~$\mu$K and 3.5~$\mu$K in the four-round-trip case. This indicates a heating of about 0.2~$\mu$K per 50~mm transport. Such a small heating rate is expected from the measured spectrum of mechanical vibrations \cite{mid11} and estimations of parametric heating \cite{sav97}.

A radial temperature of 1.4(3)~$\mu$K is determined from a time-of-flight analysis, with and without one round trip before detection. The small heating observed in the sideband spectra is not observed due to the larger uncertainty of this measurement. 
The radial temperature is also determined from sideband spectra by fitting the radial temperature, the longitudinal trap frequency and the amplitude according to the theoretical model presented in \cite{bla09a} to the data in Fig.~\ref{fig:sidebands}. A radial temperature of about 3~${\rm \mu K}$ and no significant heating is found.
The fact that the radial temperature does not match the longitudinal temperature is probably due to the long interrogation pulse, as suggested in \cite{bla09a}. 
 
These results indicate that heating due to transport is so small that it creates new opportunities for experiments with ultra cold atoms or molecules in controlled environments. In the next chapter we will focus on using this transport method in an optical lattice clock. 
%
%
%%%%%%%%%%%%%%%%%%%%%%%%%%%%%%%%%%%%%%%%%%%%%%%%%%%%%%%%%%%%%%%%%%%%%%%
%
%		Application in an optical clock
%
%%%%%%%%%%%%%%%%%%%%%%%%%%%%%%%%%%%%%%%%%%%%%%%%%%%%%%%%%%%%%%%%%%%%%%%
%
%
\section{Application in an optical clock} \label{sec:clock}
When operating a clock it is desirable to have a high duty cycle of the atom interrogation to achieve a high stability, which means that dead times as e.g. the atomic transport should be avoided or kept as short as possible. Another crucial point is the accuracy of the clock, which can be severely compromised by a residual motion of the atoms i.e. a first order Doppler effect. 

\begin{figure}[!t]
\centering
\includegraphics[width=\columnwidth]{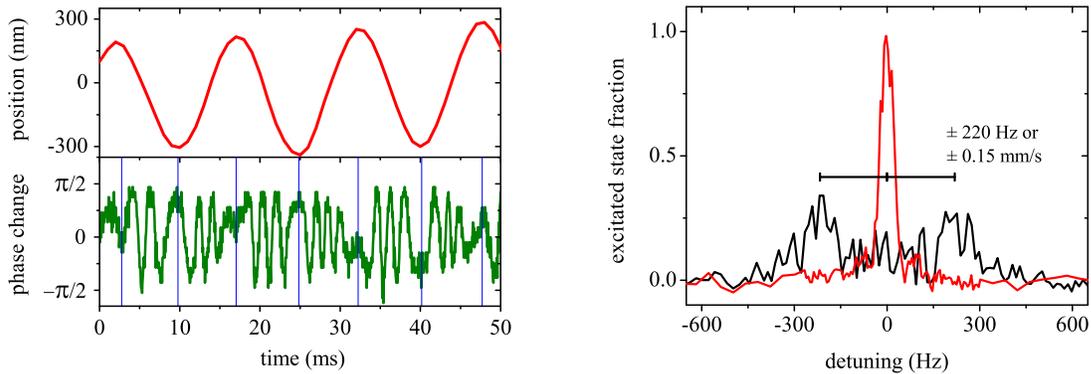}
\caption{Left: Oscillation of the reflection stage position upon arrival at the final position detected by the translation stage motion control (upper trace) and phase change of the clock laser light which is reflected by the lattice mirror (lower trace). An accumulated phase of $2\pi$ corresponds to a mirror movement of $\approx$400~nm. The blue vertical lines indicate turning points, inferred from the phase change. Right: Spectroscopy of the $^{88}$Sr clock transition without (black line) and with (red line) path length stabilisation.}
\label{fig:resmotion}
\end{figure}

We will first discuss the effect of residual motion of the lattice setup shortly after the arrival at the interrogation position. Due to the high acceleration and the large mass of the stages of about 1.5~kg each, residual motions of the optical setup in axial direction are expected. The atoms will synchronously follow the motion of the retro-reflection mirror since the propagation time which the lattice light requires to reach the atomic position is very short compared to typical mechanical oscillation periods. 

Upon arrival the translation stage motion control system indicates an axial position variation of a few 100~nm with a frequency of $\approx60$~Hz (Fig.~\ref{fig:resmotion}, left). The oscillation is damped but lasts for several 100~ms. For reasons of clock stability it is not reasonable to wait until the stages reach a sufficiently small oscillation amplitude. High resolution spectroscopy of the clock transition 
10~ms after arrival 
is however impossible with this motion of the atoms as the frequency modulation by the Doppler effect is too large (Fig.~\ref{fig:resmotion}, right). The observed frequency width of about 220~Hz corresponds to a velocity of 0.15~mm/s, which is about twice the velocity deduced from the stage oscillation.
To investigate this in more detail 
we observe the interference signal between clock laser light that is reflected by the lattice mirror and a local reference. This signal is used 
to monitor the phase change of the reflected light (Fig.~\ref{fig:resmotion}, left).
By inferring the turning points of the mirror motion from the phase change (vertical markers in Fig.~\ref{fig:resmotion}, left),
we find an oscillation frequency consistent with the built-in translation stage diagnostics. 
The oscillation amplitude of the mirror is about twice as big as that of the translation stage, 
which we attribute to acceleration related tilting of the lattice mirror. 
This tilting is consistent with the dynamic pitch discussed in Sec. \ref{sec:setup}.

\begin{figure}[!t]
\centering
\includegraphics[width=11cm]{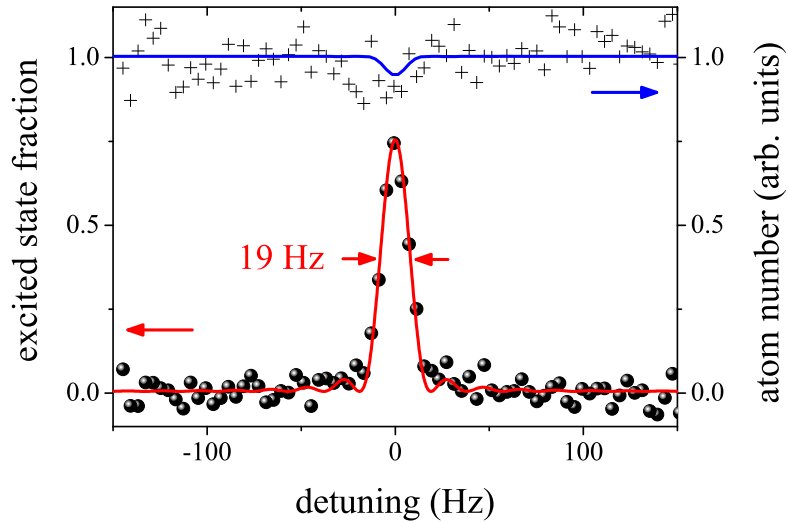}
\caption{Interrogation of the atoms, transported by 5~cm: high resolution spectrum of the clock transition (dots), total atom number (crosses) and calculated losses due to $^3$P$_0$ -- $^3$P$_0$ collisions on the way back (blue line).}
\label{fig:clock20Hz}
\end{figure}

To recover a spectroscopy signal with high contrast, it is necessary to ensure a constant phase of the clock laser field at the position of the atoms. This is achieved by stabilising the path length between the clock laser setup and the lattice mirror \cite{fal11a} that is dragging the atoms along. 
Enabling the path length stabilisation, the frequency modulation due to the Doppler effect is suppressed (Fig.~\ref{fig:resmotion}, right) and Fourier limited spectra can be recorded with the same signal contrast as without transport (Fig.~\ref{fig:clock20Hz}), indicating residual phase fluctuations of less than 1~rad amplitude. In addition, any phase drift during the pulse needs to be suppressed, which increases the demands on the length stabilisation as larger perturbations have to be removed.
By careful analysis of the residual phase errors of the stabilisation, it can be ensured that these errors will not degrade the performance of the clock.

From the investigations in Refs.~\cite{lis09,tra09} we expect collision losses when $^{88}$Sr in the $^3$P$_0$ state is trapped in the lattice. The influence of these losses is higher with transport than without because the time in which atoms in the $^3$P$_0$ state are trapped before detection is prolonged by the transport time. We have calculated the $^3$P$_0$ -- $^3$P$_0$ losses on the transport back (inter state losses are considerably smaller) and have plotted the expected atom number change as function of the clock laser detuning (Fig.~\ref{fig:clock20Hz}, blue line) together with the measured total atom number (crosses). Under our operating conditions we do not observe significant losses that might limit the operation of a lattice clock.

So far the crucial points as transport efficiency, heating of the atoms, residual motion of the setup and transport-time-enabled collisions were tackled individually. Now we want to proof that clock operation is indeed possible and how strong the stability is degraded due to transport. 
The expected fractional instability of a clock due to the signal to noise ratio $S/N$ as function of the averaging time $\tau$ is given by \cite{rie04a,tak11}
\begin{equation}
\label{eq:sigma}
\sigma_y(\tau)  = \frac{1}{K} \frac{\Delta\nu}{\nu}\frac{1}{S/N}\sqrt{t_{c}/\tau}
\end{equation}
with the clock frequency $\nu$, the FWHM linewidth $\Delta\nu$ and the cycle time $t_{c}$. The factor 
$K=\Delta\nu\cdot dS(\nu)/d\nu /S_{\rm max}$ depends on the line slope $dS(\nu)/d\nu$ at the points where the line is probed. For a Rabi-interrogation line shape, probed at the half-width points $K=1.515$. 

In a typical clock operation cycle the atoms are cooled and loaded into the lattice within 180~ms, then a ${\rm \pi}$-pulse of e.g. 80~ms duration (limited by the coherence time of the clock laser) on the half-width points of the clock transition is applied. Determining the excited state fraction takes another 40~ms. Thus we achieve an overall cycle time of $t_{c}\approx$~300~ms. For interrogation of the atoms in a specific environment away from the location of the magneto optical trap we insert two transport phases in the clock operation cycle. The atoms are loaded into the lattice, transported 50~mm away, interrogated and transported back for detection. To avoid overload of the current controller of the air bearing stages we inserted gaps of 250~ms at both positions, prolonging the cycle time to 1.3~s. After finishing the measurements we designed a trajectory that avoids overloading by a reduced maximum acceleration of $a_{\rm max}=8\:{\rm m/s^2}$. This extends the transport time by only 15~ms and enables a cycle time of 800~ms.

\begin{figure}[!t]
\centering
\includegraphics[width=9cm]{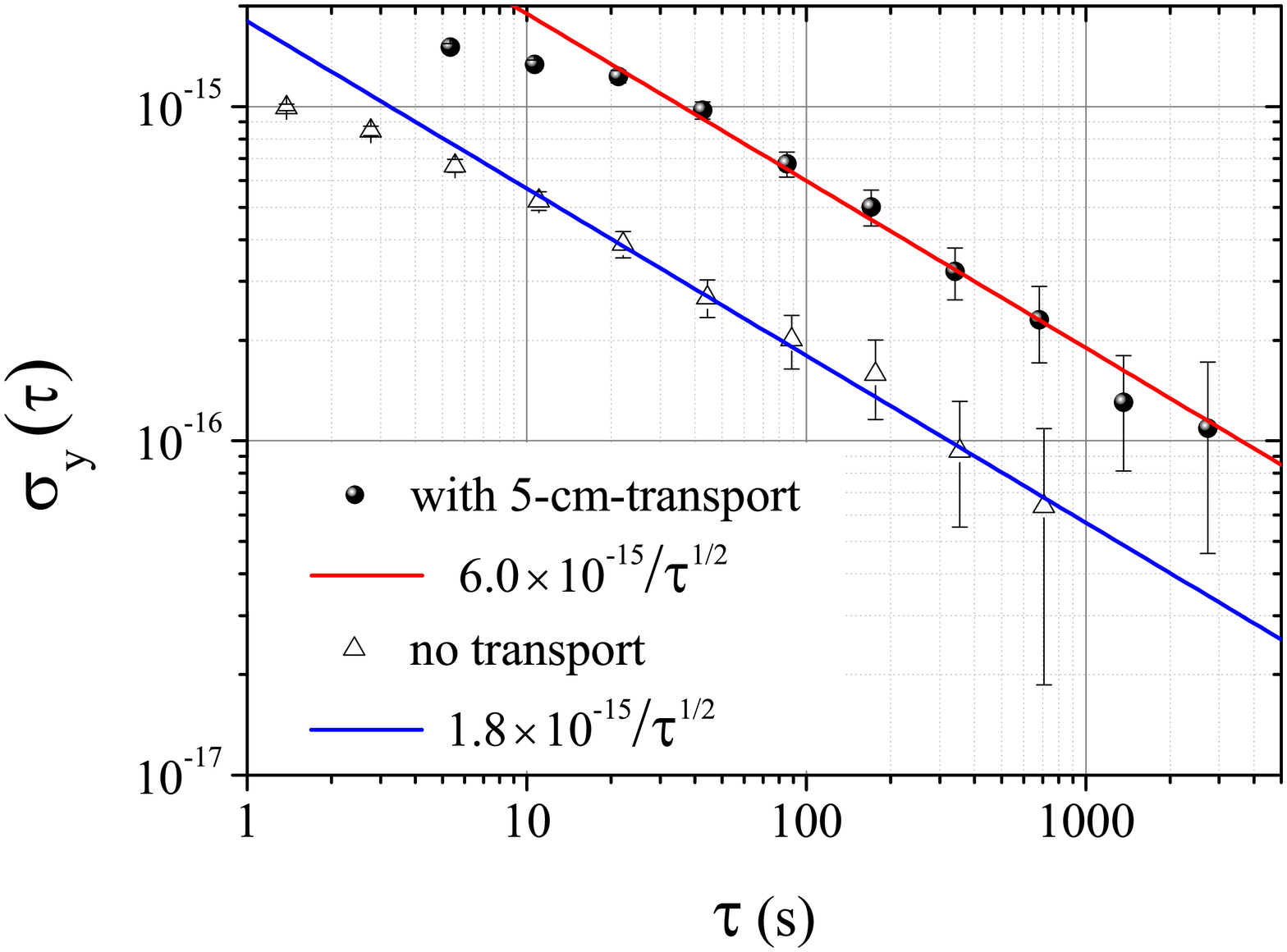}
\caption{Instability with (dots) and without (triangles) transport. In each measurement an interleaved stabilisation scheme is used, in which the clock laser is stabilised in two independent but interleaved stabilisations to the clock transition.}
\label{fig:stability}
\end{figure}

We estimate the clock stability by locking the clock laser to the reference line in two independent interleaved stabilisations \cite{deg05, deg05a}.  
From the frequency differences of the offset frequencies between the clock laser and its reference cavity in both stabilisations the Allan deviation is calculated.
Common mode frequency shifts are suppressed to a high degree.
To compare the clock stability with and without transport we recorded two individual interleaved stabilisation measurements, one with transport and one without (Fig.~\ref{fig:stability}). 
We observe a degradation of the clock stability by a factor of 3.3. A degradation by a factor of 2 due to the prolonged cycle time is expected from Eq.~\ref{eq:sigma}. The additional degradation is most likely due to the higher sensitivity to laser frequency fluctuations during the increased dead time (Dick effect) \cite{dic87, que03}.
In contrast to clock operation, in the interleaved stabilisation scheme used here twice the number of interrogation cycles is required for a single data point. 
Thus the cycle time $t_{c}$ is twice as long as if only one stabilisation is utilized.
Consequently the stability in the interleaved scheme is degraded by a factor of $\sqrt{2}\cdot\sqrt{2}$ compared to clock operation due to the doubled cycle time and the noise of two independent stabilisations influencing one data point.
The observed instability with transport of  $6\times 10^{-15}/\sqrt{\tau/{\rm s}}$ implies a clock instability of about $3\times 10^{-15}/\sqrt{\tau/{\rm s}}$, possibly better due to a reduced influence of the Dick effect.  
With the above mentioned trajectory with adjusted acceleration the fractional instability can be further improved to about $2.3\times 10^{-15}/\sqrt{\tau/{\rm s}}$.
Additional improvement is expected when lasers with smaller linewidth and instability \cite{kes12a} are applied. For Rabi-spectroscopy with an interrogation pulse duration of 1~s the resultant Fourier-limited linewidth  is 0.8~Hz. This would decrease the instability by a factor of $\approx6$ (c.f. Eq.~\ref{eq:sigma}) to $\approx4\times 10^{-16}/\sqrt{\tau/{\rm s}}$.

The presented measurements show that high performance clock operation is indeed possible with atoms transferred between preparation and interrogation regions. The transport related degradation of the stability is acceptable and investigations of systematic effects -- such as the blackbody shift -- in dedicated environments are now possible.

%
%
%%%%%%%%%%%%%%%%%%%%%%%%%%%%%%%%%%%%%%%%%%%%%%%%%%%%%%%%%%%%%%%%%%%%%%%
%
%		Conclusion
%
%%%%%%%%%%%%%%%%%%%%%%%%%%%%%%%%%%%%%%%%%%%%%%%%%%%%%%%%%%%%%%%%%%%%%%%
%
%
%
\section{Conclusion}
We have demonstrated fast transport of atoms confined in a far-detuned 1D optical lattice by 5~cm. The new transport method uses mechanical translation stages on which the optics creating the optical lattice are mounted. Our generic approach can be applied to most 1-D lattice geometries and in various applications. Accelerations of 1~g and velocities of 300~mm/s were used; both are limited by the specifications of the translation stages. The method has the potential to reach  20~nm resolution of the axial position of the trapping potential, depending on the encoder resolution.
We observe very small losses and only slight heating induced by the transport. 

When applying an optical path length stabilisation it is possible to perform spectroscopy with sub-Hertz resolution 10~ms after the transport. The degradation of the clock stability with atomic transport is acceptable and no increase of clock uncertainty is expected. The effect of reduced stability will become less severe when lasers with smaller linewidth and instability are applied \cite{kes12a} because the ratio of interrogation to preparation time will become more favourable.

The setup described here can also be of interest for other experiments that need to bring atoms in a very controlled way close to surfaces or into contact with other samples. Even collision experiments with highly controlled relative velocity can be considered. Compared to other transport options we see the advantage that in our setup the constraints on the optical lattice in terms of lattice geometry, optical power, and trap depth are small even for long transport distances.

\section{Acknowledgment}

The support by the Centre of Quantum Engineering and Space-Time Research (QUEST), the European Community's ERA-NET-Plus Programme (Grant No.~217257), and by the EU's ${\rm 7^{th}}$ framework  in the project Space Optical Clocks II (Grant No.~263500) is gratefully acknowledged.

\section*{References}
\bibliographystyle{unsrt}

\end{document}